\documentclass[a4paper]{article}
\usepackage{lipsum}
\usepackage{INTERSPEECH2021}
\usepackage{caption}
\usepackage{subcaption}
\usepackage{hyperref}
\usepackage{tabularx}

\title{WavThruVec: Latent speech representation as intermediate features for neural speech synthesis}

\name{Hubert Siuzdak$^1$\thanks{Audio samples are available at: \url{https://charactr-platform.github.io/WavThruVec/}}, Piotr Dura$^1$, Pol van Rijn$^2$, Nori Jacoby$^2$}
\address{
  $^1$Charactr, Inc.\\
  $^2$Max-Planck-Institute for Empirical Aesthetics, Frankfurt, Germany
  }
\email{huberts@charactr.com}

\begin{document}

\maketitle
\begin{abstract}
Recent advances in neural text-to-speech research have been dominated by two-stage pipelines utilizing low-level intermediate speech representation such as mel-spectrograms. However, such predetermined features are fundamentally limited, because they do not allow to exploit the full potential of a data-driven approach through learning hidden representations. For this reason, several end-to-end methods have been proposed. However, such models are harder to train and require a large number of high-quality recordings with transcriptions. Here, we propose WavThruVec -- a two-stage architecture that resolves the bottleneck by using high-dimensional \textsc{wav2vec 2.0} embeddings as intermediate speech representation. Since these hidden activations provide high-level linguistic features, they are more robust to noise. That allows us to utilize annotated speech datasets of a lower quality to train the first-stage module. At the same time, the second-stage component can be trained on large-scale untranscribed audio corpora, as \textsc{wav2vec 2.0} embeddings are already time-aligned. This results in an increased generalization capability to out-of-vocabulary words, as well as to a better generalization to unseen speakers. We show that the proposed model not only matches the quality of state-of-the-art neural models, but also presents useful properties enabling tasks like voice conversion or zero-shot synthesis.
\end{abstract}
\noindent\textbf{Index Terms}: text-to-speech, intermediate speech representation, end-to-end learning, voice conversion, zero-shot synthesis

\section{Introduction}
\begin{figure}[t]
\includegraphics[width=\linewidth]{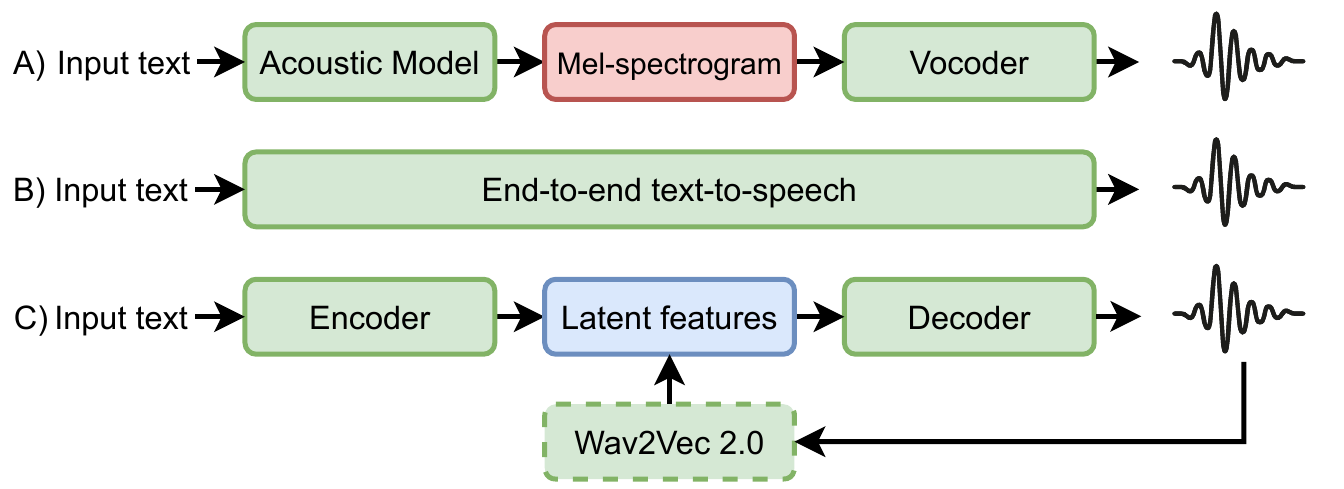}
\caption{A high-level comparison of TTS architectures: A) a traditional two-stage pipeline with mel-spectrogram as an intermediate speech representation; B) end-to-end TTS that generates waveform directly from input text; C) a proposed two-stage TTS that leverages latent speech representation from the external, pretrained model. Green blocks represent learnable neural modules, red represents predetermined features, while blue represents hidden representation. The dashed outline indicates that Wav2Vec is freezed during the training and its parameters are not updated.}
\label{fig:pipelines}
\end{figure}

The rapid development of deep neural networks has led to substantial improvements in audio quality of text-to-speech (TTS) systems. Traditionally, TTS pipelines consist of multiple components, such as text analysis, acoustic model and vocoder \cite{tan2021survey}. This modular design has been widely adopted as it decomposes the challenging problem of alignment between text or phoneme input and much longer sequences of waveform samples.
With the advances of sequence-to-sequence learning, TTS systems have been simplified to two-stage pipelines with an acoustic model that generates acoustic features directly from a text or phoneme sequence, followed by a vocoder that synthesizes a waveform (see Figure \ref{fig:pipelines}A). To bridge the components, intermediate representation in the form of low-level acoustic features was proposed, with the mel-frequency spectrogram being the most popular choice across all variety of acoustic models, such as RNN- \cite{tacotron2}, CNN- \cite{ping2018deep}, Transformer- \cite{li2019neural,ren2019fastspeech,lancucki2021fastpitch}, and Flow-based models \cite{kim2020glow}.
Mel-spectrograms became a widely accepted intermediate speech representation allowing for simultaneous development of performant neural vocoders. A breakthrough has been made with the introduction of autoregressive models for raw audio \cite{oord2016wavenet}, but their limitation in inference speed pushed researchers to investigate faster vocoders capable of parallel waveform generation such as flow- \cite{oord2018parallel, prenger2019waveglow}, GAN- \cite{donahue2018adversarial,kumar2019melgan,kong2020hifi}, or diffusion-based models \cite{chen2020wavegrad}.

Multi-stage pipelines have been criticized for being prone to carry over errors from module to module and requiring more manual labor and feature engineering (e.g. text analysis) \cite{tan2021survey}. What is more, the predetermined speech representation is often considered a bottleneck, which causes the necessity of additional finetuning \cite{sotelo2017char2wav}. The subsequent advances in neural audio synthesis have encouraged researchers to explore fully end-to-end models that discard predefined intermediate features, but directly generate waveforms conditioned on input text (see Figure \ref{fig:pipelines}B). This approach makes it possible to learn hidden, high-dimensional representations throughout the network and it has led to several promising results \cite{ren2020fastspeech,donahue2020end,weiss2021wave,kim2021conditional}.

In this work we show that by replacing low-level acoustic features with latent speech representations, we can still benefit from a practical, two-stage architecture -- such as reusing trained components in other models -- while solving problems that are typically addressed with end-to-end models. 
Specifically, we use a pretrained \textsc{wav2vec 2.0} model \cite{baevski2020wav2vec}, that has become state-of-the-art in speech recognition by learning high-level contextualized representations of speech units through self-supervision, followed by a fine-tuning procedure on annotated data. Our architecture consists of two components: An\,encoder (text2vec) which converts text input to a \textsc{wav2vec} embeddings and a decoder (vec2wav) which converts these embeddings to a waveform (see Figure \ref{fig:pipelines}C). To our knowledge, this is the first work that successfully uses a pretrained, self-supervised speech representation as intermediate acoustic representation, making TTS a kind of a downstream task.

Since \textsc{wav2vec} -- as a speech recognition model -- is supposed to produce linguistic features, the second-stage component of our TTS pipeline becomes a combination of acoustic model and vocoder, making it inherently more difficult to recover a waveform, compared to a typical vocoder conditioned on mel-spectrograms. However, such high-level intermediate representations have a wide range of benefits. The encoder (text2vec) can be trained on low-quality transcribed audio, since \textsc{wav2vec 2.0} is robust to noise. The second component (vec2wav) can be trained on a separate dataset of high-quality speech recordings without text annotations, because the latent speech features in the intermediate representation are already time-aligned (see Table \ref{tab:temps}). This can be considered as a kind of semi-supervision, since only part of the training data needs to be annotated, solving a recurrent issue in TTS that recordings should be both of high quality and contain transcripts.

In the present paper, we show that training the decoder on large untranscribed speech datasets, containing an order of magnitude more speakers than typical TTS dataset, (i) leads to strong generalization capability to unseen speakers, enabling zero-shot synthesis.  At the same time, the encoder can take advantage of the widely available, varying audio quality, transcribed datasets in order to implicitly learn pronunciation, relieving the need for external grapheme-to-phoneme models, resulting in (ii) better generalization to out-of-vocabulary words. Furthermore, we show that \textsc{wav2vec} embeddings -- in contrast to acoustic features such as mel-spectrograms -- are speaker-independent, (iii) enabling voice conversion by design.

\begin{table}[]
\caption{Data requirements for different architectures. HQ stands for high-quality audio, which is determined by a number of factors such as recording equipment or background noise. We relax this requirement in our encoder (text2vec).}
\begin{tabular}{c|cc|c|cc}
          \multicolumn{1}{c}{}   & \multicolumn{2}{c}{Two-stage TTS} & \multicolumn{1}{c}{E2E}            & \multicolumn{2}{c}{WavThruVec} \\
            & Model      & Vocoder     & Model & Enc.       & Dec.      \\
\hline
HQ audio    & +                   & +           & +              & --              & +             \\
Transcribed & +                   & --           & +              & +              & --           
\end{tabular}

     \label{tab:temps}
\end{table}

\section{Related work}

As we discard acoustic-level intermediate features in favor of high-level, contextualized representations of speech units, our approach can be compared to TTS methods that generate audio waveforms directly from linguistic features. WaveNet \cite{oord2016wavenet} was introduced as such a model, although it is now more commonly employed as a vocoder that converts mel-spectrograms into audio waveforms. GAN-TTS \cite{binkowski2019high} also utilizes a sequence of linguistic features as an input to the model. However, such linguistic features are derived from text-analysis modules that typically require hand-crafted annotations such as phonemes, syllables, durations, stress or intonation. To this end, we utilize hidden activations of the model that learnt the structure of speech without such supervision.

Self-supervised speech representations have been recently demonstrated to be useful for the task of resynthesis and voice conversion \cite{polyak21_interspeech}, however the text-to-speech problem hasn't been addressed. The subsequent work \cite{10.1162/tacl_a_00430} further investigates self-supervised speech representations in the context of generating spoken language, but employs abstract pseudo-text features to provide linguistic conditioning. Inspired by these works we propose a downstream task of applying these high-level speech representations for text-to-speech. 

\begin{figure}[t]
    \centering
     \includegraphics[width=\linewidth]{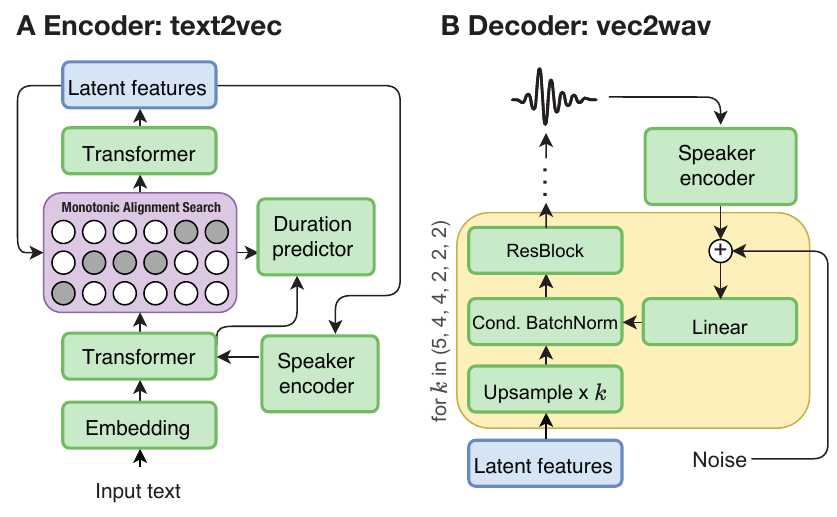}
    \caption{Our architecture consists of an encoder (text2vec) and a decoder (vec2wav) with separate speaker encoders that learn the corresponding speaker embeddings.}
    \label{fig:figure}
\end{figure}

\section{Method}

As in the typical two-stage TTS pipeline, the proposed architecture is composed of two neural networks that can be trained independently (Figure \ref{fig:figure}). To bridge the components we use hidden activations of \textsc{wav2vec 2.0} \cite{baevski2020wav2vec} for a particular audio sample. Specifically, we use pretrained checkpoint of Wav2Vec 2.0 Base model (with 768 latent dimension), finetuned on 960 hours of LibriSpeech. Although the presented two-stage architecture with latent, learnt intermediate representation is a high-level design and can be realized with a variety of neural submodules, we introduce WavThruVec -- a baseline model consisting of Transformer-based text2vec and GAN-based vec2wav.

\subsection{text2vec} 
The first-stage component of our pipeline mostly follows the FastSpeech \cite{ren2019fastspeech} architecture with two blocks of Feed-Forward Transformers (FFT) consisting of a self-attention and 1D convolutional network (Figure \ref{fig:figure}A). Instead of using a teacher-based length regulator between the FFT blocks as in the original work, we utilize unsupervised Monotonic Alignment Search introduced by \cite{kim2020glow}. We specifically train soft and hard alignments with additional diagonal prior as in \cite{shih2021radtts}. The soft alignment matrix $\mathcal{A}_{soft} \in \mathbb{R}^{N \times T}$ is based on the learned pairwise affinity between all text tokens $\phi \in \Phi$ and \textsc{wav2vec 2.0} activations $x \in X$ of lengths $N$ and $T$ respectively. The forward-backward algorithm is used  to maximize the likelihood $P\left(s_{t}=\phi \mid x_{t}\right)$, where $s_t$ is a random variable for a text token aligned at timestep $t$ with target $x_t$. To obtain a binary alignment map $\mathcal{A}_{hard}$, the Viterbi algorithm is used, and to further close the gap between soft and hard distributions, their KL-divergence is minimized: $\mathcal{L}_{bin}=\mathcal{A}_{hard} \odot \log \mathcal{A}_{soft}$. Hard alignment serves as a target for the duration predictor that is trained via Mean Squared Error loss (MSE) to be used at inference time. Similarly, the model optimizes MSE between predicted and target speech representation. For the multi-speaker setup, we condition the first FFT block on the speaker embedding that is obtained through feeding the target sequence into a series of convolution layers followed by channel-dependent frame attention pooling \cite{desplanques2020ecapa}. Such an encoder is supposed to capture the style of a particular speaker with regards to some prosody features that are represented in \textsc{wav2vec 2.0} latent variables. It can be used at inference time to produce speaker embedding in a zero-shot manner.

\subsection{vec2wav} 
The role of the second-stage component is to generate an audio waveform conditioned on hidden activations of \textsc{wav2vec 2.0} (Figure \ref{fig:figure}B). 
Vec2wav is a Generative Adversarial Network, based on the HiFi-GAN \cite{kong2020hifi}, consisting of a fully convolutional generator and several sub-discriminators. The generator upsamples input features through the sequence of transposed convolutions followed by residual blocks of dilated convolutions. Similarly to \cite{binkowski2019high}, we introduce Conditional Batch Normalization to condition the network on the speaker embedding between the residual blocks at different temporal resolutions. Each Conditional Batch Normalization is preceded by a linear network that takes the speaker embedding concatenated with a vector of random numbers from a normal distribution. We synthesize speech at a sampling rate of 32 kHz while our input features have temporal resolution of 50 Hz, resulting in 640x upsampling factor, compared to 256x of original HiFi-GAN. Therefore the configuration of the generator was changed for upsample rates to a sequence of (5, 4, 4, 2, 2, 2) with corresponding kernel sizes (11, 8, 8, 4, 4, 4), while the hyper-parameters of residual blocks are the same as in HiFi-GAN V1. Additional multi-period sub-discriminators are added with periods of $[13, 17, 19]$ to obtain the receptive field of similar length. To enable multi-speaker capabilities, we do not use learnable embeddings through a look-up table, but rather train a speaker encoder that takes mel-spectrogram of a particular sample as an input and produces fixed-length embedding. Specifically ECAPA-TDNN \cite{desplanques2020ecapa} architecture is used as a speaker encoder.

\section{Experiments}
\subsection{Datasets} 

As the targets of the first-stage module are noise-robust, we can train on speech corpora with varying audio quality, typically designed for automatic speech recognition. We can use audio recordings sampled at 16 Khz, since \textsc{wav2vec 2.0} has learnt speech representation on such data. In particular, we use LibriSpeech \cite{librispeech} and CommonVoice \cite{commonvoice} English datasets, comprising a total of about 3,000 hours of annotated speech recordings. We do not use any rule-based text normalization or phonemization methods, but train the model on raw character inputs.

Vec2wav, on the other hand, can be trained without text transcriptions, but instead requires higher quality speech data, as it affects the overall naturalness of generated samples. We therefore use AVSpeech \cite{ephrat2018looking}, a large-scale audio-visual dataset comprising speech video clips with no interfering background noises, initially intended for speech separation and audio-visual event localization. We collected only audio recordings in English and performed the following preprocessing steps: (1) each sample was processed using the NeMo toolkit \cite{kuchaiev2019nemo} for voice activity detection to discard unspoken fragments, (2) followed by speaker diarization to ensure there is a single speaking person per clip; (3) a pretrained neural model for sound event detection \cite{kong2020panns} was used to filter out the recordings with much background sound, and (4) the audio tracks were downsampled to 32 kHz. As a result, we obtained high quality recordings of 11,876 distinct speakers. We additionally use Hi-Fi TTS \cite{bakhturina2021hi} and VCTK \cite{yamagishi2019cstr} datasets. For the finetuning stage of both text2vec and vec2wav models, only VCTK dataset is used.

\subsection{Training} 
Text2vec is trained using the LAMB optimizer with learning rate of 0.1, $\beta_{1}=0.9$, $\beta_{2}=0.98$, $\epsilon = 10^{-9}$, similarly to \cite{lancucki2021fastpitch}. We follow the training schedule of \cite{shih2021radtts}, to add the binarization term and hard alignments to the loss function. 

The discriminators and the generator of the GAN-based vec2wav are trained as in \cite{kong2020hifi}, using the AdamW optimizer with $\beta_{1}=0.8$, $\beta_{2}=0.99$, weight decay $\lambda=0.01$ and learning rate decaying by a 0.999 factor in every epoch with an initial value of $2\cdot10^{-4}$. Our intermediate representation is already aligned, so we do not have to incorporate dynamic time warping to relax alignment constraint in the spectrogram prediction loss as in \cite{donahue2020end}. However, we linearly decay its weight coefficient to make the loss function increasingly dependent on the GAN objective. Similarly to \cite{ren2020fastspeech, donahue2020end}, we adopt the windowed generator training with a training window of 0,64\,s.

Both text2vec and vec2wav were trained on 4 NVIDIA V100 GPUs with batch sizes of 32 and 24, respectively. After 800k iterations of pretraining, both models are finetuned for 80k iterations on VCTK dataset with a 10-fold lower value of initial learning rate.

\subsection{Experimental Setup for Comparison}

We compare our model with the following implementations: Tacotron 2\footnote{https://github.com/NVIDIA/tacotron2}, Fastpitch\footnote{https://github.com/NVIDIA/DeepLearningExamples} and VITS\footnote{https://github.com/jaywalnut310/vits}. Since Tacotron 2 and Fastpitch repositories don’t come with pretrained checkpoints on VCTK datasets, we train our own up to 500k iterations on 4 NVIDIA V100 GPUs. Similarly to \cite{kim2021conditional}, for both Tacotron 2 and Faspitch, we use HiFi-GAN as a second-stage model, and further finetune it for 200k iterations with Tacotron 2 outputs via teacher-forcing.

We conduct crowd-sourced listening tests to (i) evaluate the overall quality of the samples and (ii) evaluate the generalization capability to unseen words. 36 US participants (17 female, 19 male) took part in the experiment. The age ranged from 19 to 63 years old (\textit{M} = 36, \textit{SD} = 11). All participants were recruited from Amazon Mechanical Turk (AMT) with requirement of minimal age of 18 years, 99\% or higher approval rate on at least 5,000 previous tasks on AMT, residency in the US and wearing headphones (had to pass a pre-screening headphone check \cite{woods2017headphone}). Participants  provided informed consent in accordance with the Max Planck Society Ethics Council approved protocol (application 2021\_42). For the evaluation, all audio samples are normalized to -3 dBFS, resampled to 22,05 kHz, and the beginning and trailing silence is trimmed. Raters were allowed to evaluate each audio clip once.

\section{Results}
\subsection{Speech Synthesis Quality}

To evaluate the overall speech synthesis quality we collect Mean Opinion Score (MOS) for all the neural models outputs as well as ground truth recordings. For a random sample from VCTK dataset, participants were asked to rate the naturalness of the speech on a 5 point scale from 1 to 5. The results show that WavThruVec outperforms other TTS systems, with little difference from the ground truth score (Table \ref{tab:scores}).
Interestingly, the objective measure of similarity between the generated voices and the real ones, shows that WavThruVec performs the worst among the compared models. We hypothesize that further finetuning on VCTK dataset would lead to better similarity score, however sacrificing the generalization capability to unseen voices, that we demonstrate in section \ref{zeroshot}.

\begin{figure}[t]
     \centering
     \begin{subfigure}[b]{0.49\linewidth}
         \centering
         \includegraphics[width=\textwidth]{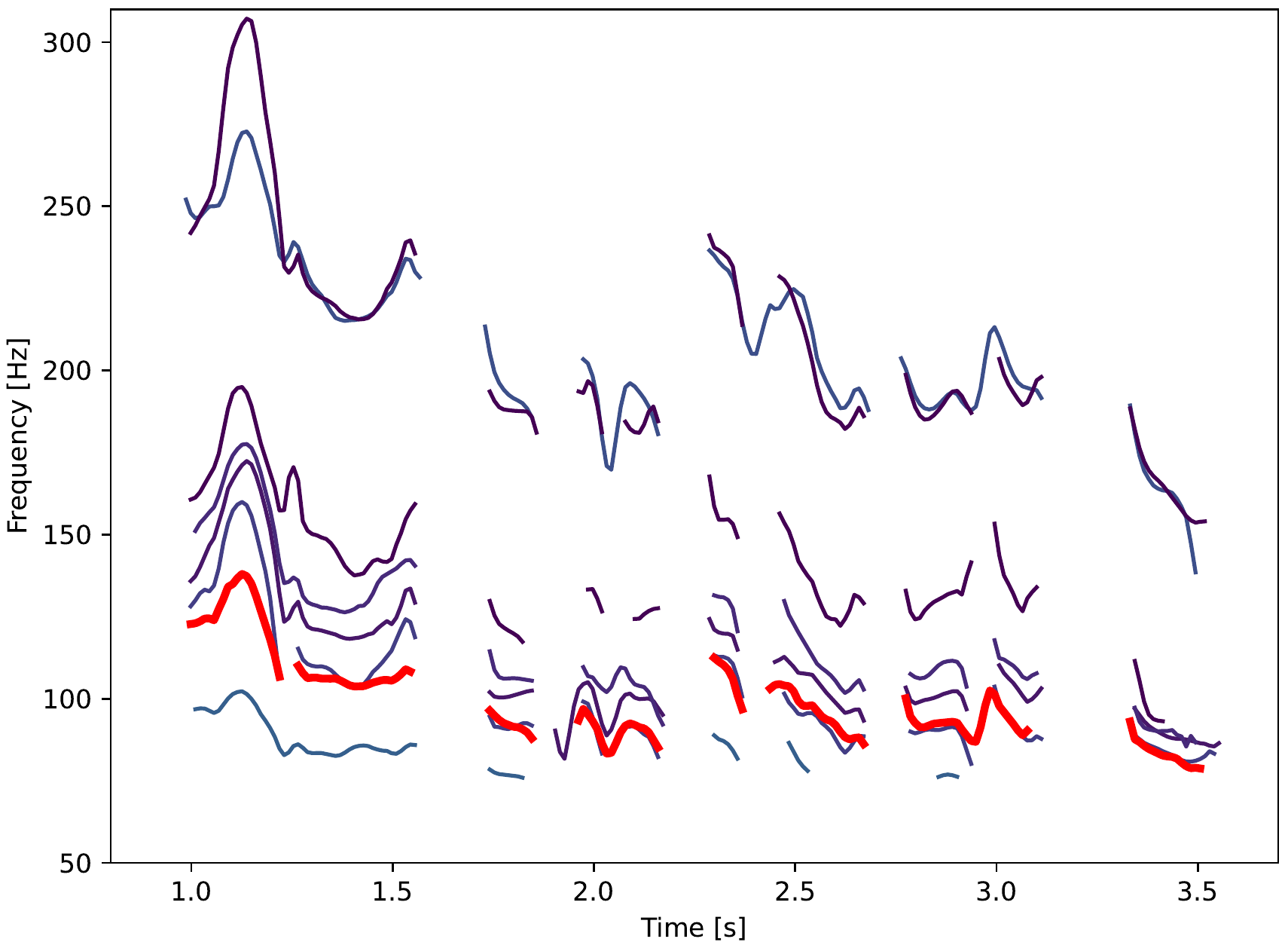}
         \caption{VITS}
         \label{fig:vc-vits}
     \end{subfigure}
     \begin{subfigure}[b]{0.49\linewidth}
         \centering
         \includegraphics[width=\textwidth]{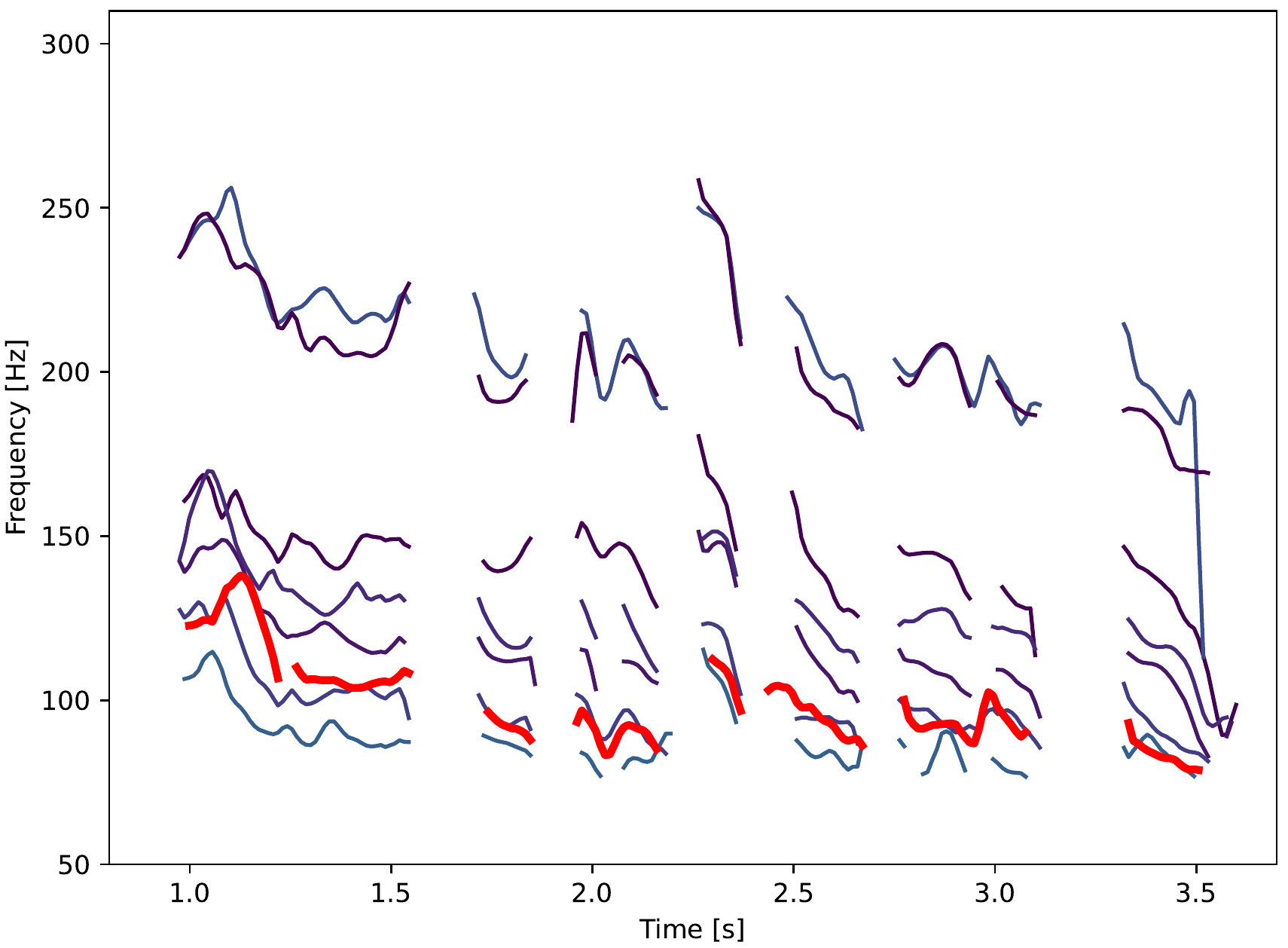}
         \caption{WavThruVec}
         \label{fig:vc-wtv}
     \end{subfigure}
        \caption{Voice conversion: the trajectories of fundamental frequency for particular speakers. The red bold line indicates the input utterance, while the other lines represent converted voices.}
        \label{fig:pitch}
\end{figure}

\begin{table}
\centering
\caption{Comparison of evaluated MOS and pronunciation errors (\% correct) with 95\% confidence intervals}
\label{tab:scores}
\begin{tabularx}{\linewidth}{X X X}
\hline
Model & MOS (CI) & \% correct (CI)\\
\hline
Ground Truth & 4.17 (±0.10) & -- \\
Tacotron 2 & 3.92 (±0.13) & 0.78 (±0.05)\\
FastPitch & 3.67 (±0.12) & 0.82 (±0.05)\\
VITS & 3.99 (±0.12) & 0.86 (±0.05)\\
WavThruVec & \textbf{4.09} (±\textbf{0.10}) & \textbf{0.89} (±\textbf{0.04})\\
\hline
\end{tabularx}
\end{table}

\subsection{Generalization to Unseen Words}
To prepare an out-of-vocabulary list, we took a collection of English words (384k)\footnote{Wordlist 3, https://www.keithv.com/software/wlist/} and removed those that were seen in any of the training sets. Then by querying SKELL\footnote{https://skell.sketchengine.eu/} with random words from the list, we collected valid English sentences and synthesized them with the neural models. Since these words are out of the open-sourced pronouncing dictionaries, we conduct crowd-sourced listening tests to evaluate pronunciation errors. The results showing that WavThruVec outperforms other TTS systems are presented in Table \ref{tab:scores}.

\begin{table}[t]
\caption{Averaged objective measure of similarity between generated outputs and ground truth samples. Values represent cosine similarity between speaker embeddings from an external speaker verification model \cite{desplanques2020ecapa}. For zero-shot TTS we evaluate different lengths of a target-speaker sample.}
\label{tab:similarity}
\centering
\begin{tabular}[t]{c c c c c}
\hline
& Tacotron 2 & Fastpitch & VITS & WavThruVec\\
\hline
\multicolumn{5}{c}{\textit{Text-to-speech}} \\
 & 0.65 & 0.64 & \textbf{0.69} & 0.61 \\
\hline
\multicolumn{5}{c}{\textit{Voice conversion}} \\
 & -- & -- & 0.62 & \textbf{0.64} \\
\hline
\multicolumn{5}{c}{\textit{Zero-shot text-to-speech}} \\
1 s & -- & -- & -- & 0.29 \\
3 s & -- & -- & -- & 0.43 \\
10 s & -- & -- & -- & 0.43 \\
30 s & -- & -- & -- & \textbf{0.44} \\
\hline
\end{tabular}
\end{table}

\subsection{Voice conversion}

We can bypass the text encoder (text2vec) and directly provide linguistic features for the decoder (vec2wav) by feeding a particular speech sample through \textsc{wav2vec 2.0}. Experimentally, we found out that such features can be then synthesized by vec2wav, conditioned on different speaker embeddings. It results in different voice characteristics, corresponding to the particular speakers. This implies that intermediate features we use in our pipeline are speaker-independent, enabling high-fidelity voice conversion.
The trajectory of fundamental frequency for converted voices shows that the prosody of an input utterance is substantially preserved. However, individual phonetic segments can vary between speakers (Figure \ref{fig:vc-wtv}), while VITS forces the converted voice to match the input one’s characteristics (Figure \ref{fig:vc-vits}). This is particularly evident, for example, in the case of the source speaker’s speech impediment, which will be preserved throughout all the converted voices. In contrast, WavThruVec discards most of the acoustic properties through intermediate representation and recovers a waveform highly similar to the target speaker. This is confirmed by the objective similarity results in Table \ref{tab:similarity}.

\subsection{Zero-shot text-to-speech} \label{zeroshot}

WavThruVec is not constrained to synthesize voices only from the training set and it can easily obtain target speaker embeddings given a particular speech sample. Moreover, we used an unprecedented number of training voices, as we were not limited to the transcribed recordings. To evaluate whether it led to increased generalization capability to out-of-training speakers, we use the DAPS \cite{daps} as a test set of 10 male and 10 female unseen voices. To calculate the speaker embeddings, we randomly take a variable-length sample from reference recording, and pass it through the corresponding speaker encoders. Then they serve as conditioning for both text2vec and vec2wav modules. We show that with only 3 seconds of a reference recording, WavThruVec is able to faithfully reproduce an out-of-sample voice, however there is still a large gap between similarity score of seen and unseen voices (Table \ref{tab:similarity}). Although the objective similarity score is not substantially higher for the reference recordings longer than 3 seconds, we internally found out that embeddings calculated from 10 and 30 seconds long samples resulted in higher subjective quality.

\section{Conclusions}

In this work, we demonstrate that by using latent intermediate speech representation instead of predetermined features, we combine the advantages of end-to-end learning with the practicality of a two-stage pipeline. 
This novel procedure also introduce weaker data requirements, which allows us to leverage additional speech datasets, resulting in less pronunciation errors and zero-shot synthesis capabilities. 
Most importantly, WavThruVec receives the highest scores in the crowd-sourced listening tests with regards to speech naturalness, outperforming state-of-the-art TTS systems. Although we use WavThruVec as a two-stage architecture, future work can explore joint training with a \textsc{wav2vec}-like objective. Extending this approach with additional modules e.g. for prosody modeling, may lead to further improvements. Taken together, our results show the power of latent speech representation for deep generative modeling.

\clearpage
\bibliographystyle{IEEEtran}

\bibliography{mybib}

\end{document}